\documentclass[prl,showpacs,aps,twocolumn]{revtex4}
\usepackage{amsmath}
\usepackage{epsfig}
\begin{document}

\bigskip
\title{No quasi-long-range order  in the two-dimensional liquid crystal}

\author
{
Ricardo Paredes V.$^{\dagger}$,  Ana Isabel Fari\~nas-S\'anchez$^{\ddagger}$
and
Robert Botet$^{\star}$
 }
 
\affiliation{
$^{\dagger}$ Instituto Venezolano de Investigaciones Cient\'{\i}ficas, Centro de F\'{\i}sica, 
Laboratorio de Fisica Estad{\'\i}stica Apdo.  20632, Caracas 1020A, Venezuela \\
$^{\ddagger}$ Universidad  Sim\'on Bol{\'\i}var, Dept Fis, Apdo. 89000, Caracas 1080A, Venezuela\\
$^{\star}$
Laboratoire de Physique des Solides B\^{a}t.510, CNRS UMR8502 - Universit\'{e} Paris-Sud,
F-91405 Orsay, France  
}
 
\date{\today}
 
\begin{abstract}
\parbox{14cm}{\rm 
Systems with global symmetry group $O(2)$ experience topological transition in the 2-dimensional space. 
But there is  controversy about such a transition for systems with global symmetry group $O(3)$. 
In this paper, we study the Lebwohl-Lasher model for the two-dimensional liquid crystal,
using three different methods independent of the proper values of possible  critical exponents.
Namely, we analyze the at-equilibrium order parameter distribution function 
with: 1) the hyperscaling relation; 2) the first scaling collapse for the probability distribution function;  
and 3) the Binder's cumulant. We give strong evidences for definite lack of a line of critical points at low temperatures in the Lebwohl-Lasher model, contrary to conclusions of a number of previous numerical studies. 
}
 
\end{abstract}
\bigskip 
\maketitle

{\it Introduction.} 
Mermin and Wagner \cite{MerWag} established that no ferromagnetic phase nor any long range order 
can appear for systems of continuous symmetry at finite temperature in space dimension $d \le 2$.
However, such systems might have another type of transition governed by binding-unbinding topological defects 
at definite positive temperature $T_{\mbox{\tiny{BKT}}}$\cite{Bere,KT,KosterlitzH}. 
This kind  of topological phase transition 
is called  Berezinskii, Kosterlitz and Thouless (BKT)  transition.

The two-dimensional (2d) XY-model, with global symmetry group $O(2)$, 
exhibits such topological transition\cite{KosterlitzH}.
Quasi-long-range order (QLRO) appears at low temperatures $T$, and the order parameter vanishes 
as a power law at the thermodynamic limit. 
Due to the QLRO behavior, the system susceptibility $\chi$, that measures the fluctuations of the order parameter, 
diverges for all temperatures $T\le T_{\mbox{\tiny BKT}}$, 
and the system is characterized by a line of critical points 
below the critical temperature $T_{\mbox{\tiny BKT}}$. 
Close to $T=0$, correlations are dominated by spin-wave solution: in the 
units system where $k_{\mbox{\tiny B}} =1$ and the coupling factor 
between magnetic moments is $J=1$, the correlation function exponent, $\eta$, depends on the 
temperature as: $\eta = T/2\pi$. 
Another characteristic behavior of this transition is that at temperatures just above the BKT transition,
$t=(T-T_{\mbox{\tiny BKT}})/T_{\mbox{\tiny BKT}} \gtrsim 0$, the correlation length, $\xi$, 
diverges  as the essential singularity: $\xi \sim \exp(bt^{-1/2})$, that is much strongly than the 
ordinary second order transition power law, $\xi\sim t^{-\nu}$. 

On the other hand, Polyakov \cite{Poliakov}, using renormalization group theory,
proved that the 2d Heisenberg model, with global symmetry group $O(3)$, does not present 
any sort of phase transition.
An important difference between the two systems above is that the
global symmetry group for the XY-model is abelian, while it is not for the Heisenberg model. 
However, things are not as clear:
numerical evidences were recently given for transition \cite{Tesis, Niedemayer,PatrasciouSeiler,Patrasciou, Aguado} 
in this system, and a possible QLRO phase \cite{Tesis, Kapikranian} at very low temperatures. 
In the same spirit, it has also been reported that the 2d fully frustrated anti-ferromagnetic Heisenberg model 
presents a crossover produced by the binding-unbinding of topological defects in a very narrow temperature interval. 
In this case no  QLRO behavior at or below the transition \cite{Kawamura, Wintel} has been observed.
At present, this controversy on systems with continuous non-abelian symmetry is not solved. 

Kunz and Zumbach (KZ) \cite{KZ} performed intensive Monte Carlo study of the 2d $RP^2$ model,
which has the global symmetry group $O(3)$ and the local symmetry group $Z_2$. 
The model describes the isotropic-nematic transition of a liquid crystal.
KZ concluded with BKT-like transition from analysis of energy, specific heat and topological quantities.
But the correlation length behavior for $t \gtrsim 0$ was not proven to be either of the power-law or of the 
essential singularity type.  
Ten years later, the problem of phase transitions for liquid crystals in $d=2$ was 
complemented \cite{Tesis,PLA,RMex} using the powerful techniques of conformal transformations \cite{Cardy} (CT) 
and finite size scaling (FSS) \cite{Tesis, RMex}. 
The Lebwohl-Lasher \cite{LebLash} (LL)  was preferred to the $RP^2$ model
though sharing the same symmetries. These studies concluded with a BKT-like transition
and a QLRO phase below the BKT temperature estimated by KZ \cite{KZ} to be $T_{\mbox{\tiny BKT}}=0.513$.
At low temperatures, a spin wave dependence $\eta \propto T$ was obtained. 
To support this conclusion Dutta and Roy \cite{Dutta} 
showed that the transition is driven by topological stable points defects 
known as $\frac{1}{2}$-disclination points. Using the system susceptibility, $\chi$,
FSS was able to estimate the value of the correlation function exponent $\eta$ within
the temperature range $T \le T_{\mbox{\tiny BKT}}$. On a line of critical points, $\chi$ should scale 
with the exponent $\gamma/\nu$, which is related to $\eta$ through the hyperscaling law:
\begin{equation}
\label{hyperscaling}
\gamma/\nu = 2 - \eta.
\end{equation}
Using (\ref{hyperscaling}), estimation of the values of $\eta$ was performed \cite{Tesis,FarParBot}.
The values appeared to behave similarly to the ones obtained through CT and scaling of the order parameter, 
but there is a discrepancy of about $5\%$ between both results. 
The origin of such difference was tentatively explained arguing that the system sizes were far from 
the thermodynamic limit and the number of independent realizations were too small to reach good statistics.

The purpose of this article is to revisit the problem of the possible appearance of
BKT-like transition for the 2d LL model. 
In this model, liquid-crystal molecules are represented by unitary vectors $\vec{\sigma}_i$ 
situated on the sites, labelled $i$, of a hypercubic lattice $\Lambda$ of length $L$. 
The Hamiltonian is given by:
\begin{equation}
-\beta H = \sum_i\sum_\delta P_2(\vec{\sigma}_i\cdot \vec{\sigma}_{i+\delta}), \label{Ham}
\end{equation}
where $\beta = 1/T$,  $P_2$ is the second Legendre polynomial and the interaction is between nearest neighbors. 
The appearance of the $P_2$ function in (\ref{Ham}) comes from the $Z_2$ local symmetry. 
In the nematic phase, the preferential direction is characterized by the unit vector ${\bf n}$, 
called the director, and one can measure the local orientation with respect to the director 
by:  ${\vec \sigma}_i \cdot{\bf n} = \cos \theta_i$. 
Then, the local order parameter is defined by 
$m(i) = \langle P_2( \cos \theta_i ) \rangle$. 
Whenever the system is completely ordered: $m(i) = 1$. 

Therefore, we start from the hypothesis that the 2d LL model experiments 
BKT transition at $T_{\mbox{\tiny BKT}} = 0.513$, 
similarly to the transition observed in the 2d XY-model at $T_{\mbox{\tiny BKT}}=0.893$. 
Below such critical temperature a line of critical points should be observed in both models. 
To validate this point, we performed Monte Carlo simulations using the Wolff algorithm \cite{Wolff} in $d=2$, 
with periodic boundary conditions at temperatures well below $T_{\mbox{\tiny BKT}}$. 
A total of $6\times10^6$ independent realizations were performed for each system size and  each temperature 
for both models. Then we found  estimates of the order parameter probability distribution function (PDF), 
and estimation of the validity for the hyperscaling relation (\ref{hyperscaling}). 
Finally we will analyze the Binder's cumulant behavior, comparing also with the Heisenberg model. \\

{\it Hyperscaling relation check. --}
For the XY-model at $T=0.6$, 
we observe that both the order parameter and the susceptibility  have power law behavior, 
$\langle m\rangle \sim L^{-\beta/\nu}$ and $\chi \sim L^{\gamma/\nu}$ respectively. 
For the XY system the exponents obtained were $\beta/\nu = \eta/2  \approx 0.058$ and $\gamma/\nu \approx 1.877$. 
With use of the CT method, Berche et al \cite{EPL} obtained the value $\beta/\nu = 0.0595$ in excellent agreement 
with our results. Hyperscaling relation (\ref{hyperscaling}) is satisfied with error smaller than $0.4\%$. 

For the LL model at $T=0.4$, we obtained again excellent power laws for $\langle m \rangle$ and $\chi$ 
with respective exponents $\beta/\nu = \eta/2  \approx 0.0945$ and $\gamma/\nu \approx 1.868$. 
But now, the agreement for Eq.(\ref{hyperscaling}) is poor and about $3\%$ 
(one order of magnitude larger than for the XY case). 
The actual increase of the number of independent realizations does not really improve  the 
results obtained previously \cite{Tesis,FarParBot}. 

\begin{figure}[h]
\includegraphics[width=8 cm]{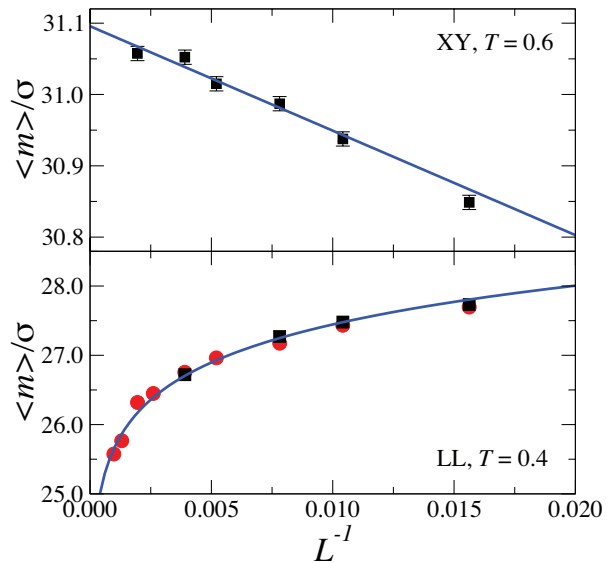}
\caption{\label{Fig1} ({\it color online}) $\langle m \rangle/\sigma$ is plotted {\it vs} $L^{-1}$ for the XY-model at  $T=0.6$ (top) and 
for the LL-model at $T=0.4$ (bottom). A linear fit is obtained for the XY-model ($\langle m \rangle/\sigma = 31.1 -16.4/L$). A power law fit shows that no saturation is observed for the LL-model. Both fits are shown as bold lines. The bold circles are the data from \cite{Tesis}. The number of independent realizations used  to obtain the bold squares is almost two orders of magnitude larger  than in \cite{Tesis}. The hyperscaling relation (\ref {hyperscaling}) is {\em not} satisfied in the thermodynamic limit by the LL-model at $T=0.4$.}
\end{figure}

We shall use now alternative method to check the hyperscaling relation. 
Let us introduce $\sigma$ as the standard deviation of the order parameter. 
One has : $\sigma^2 \propto \chi/L^{d}$, then $\sigma$ scales with 
the system size as: $\sigma \sim L^{\gamma/2\nu-1}$
for the 2d systems. 
Therefore, the ratio 
$\langle m \rangle/\sigma$ should be a constant whenever the hyperscaling relation 
(\ref{hyperscaling}) is satisfied. The great advantage for using this ratio is that 
previous estimation of the exponents is not necessary to check (\ref{hyperscaling}). 
In Fig. \ref{Fig1}, $\langle m \rangle/\sigma$  
is plotted versus $L^{-1}$ for the XY (above) and the LL (below) models 
in the low temperature domain. 

For the XY-model, the ratio is seen to saturate at the thermodynamic limit
to a value $\langle m \rangle/\sigma \simeq 31.1$. 

For the LL-model, power law is the best fit consistent with our data for $\langle m \rangle/\sigma$ versus $L$. 
The ratio does not saturate to a finite value and we conclude that the hyperscaling relation 
(\ref{hyperscaling}) does not hold in this case. 

Similar behavior was observed for the XY-model at $T_{\mbox{\tiny BKT}}=0.893$ \cite{RicBot} and 
for the LL-model at $T_{\mbox{\tiny BKT}}=0.513$ \cite{FarParBot}. 

\begin{figure}[h]
\includegraphics[width=8 cm]{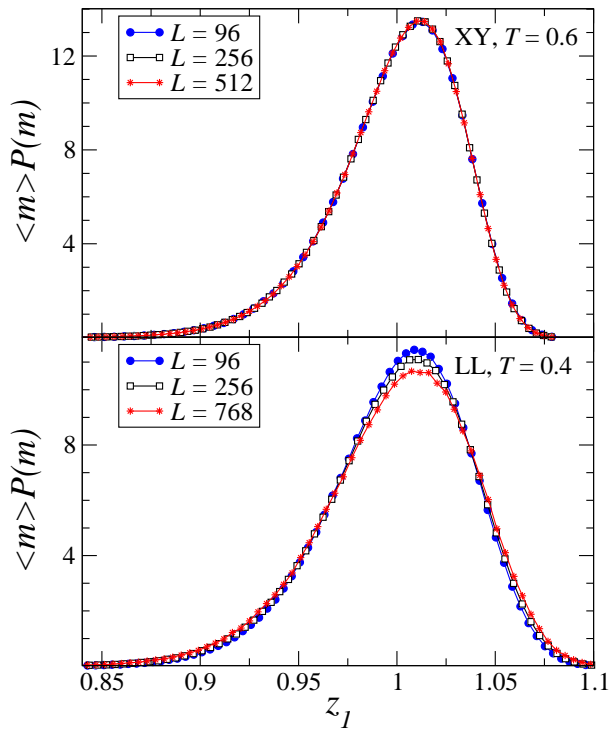}
\caption{\label{Fig2} ({\it color online}) Order parameter PDF for the XY-model at $T=0.6$ (top) and the LL model at $T=0.4$ (bottom) 
in the first scaling form. A perfect collapse is observed for the XY-model. This is not the case for the LL model. The $L=768$ data are from \cite{Tesis}, with $9\times10^4$ independent realizations. It is clear from this figure that the number of independent realizations used in \cite{Tesis} was large enough to realize the first-scaling law. {\em No} self-similarity is observed for the LL model at $T=0.4$.}
\end{figure}

{\it First-scaling relation check. --} The first-scaling law \cite{Botet1}:
\begin{equation}
\label{fS}
\left <m \right> P(m) = \Phi_T(z_1), \qquad \mbox{with} \quad  z_1 \equiv \frac{m}{<m>},
\end{equation}
should be satisfied anywhere on the line of critical points below the BKT transition.
In (\ref{fS}), $P(m)$ denotes the order parameter PDF. 
The scaling function $\Phi_T$ depends only on the actual temperature.
Here too, one great advantage of the first-scaling law is that Eq.(\ref{fS}) does not require  
knowledge of any critical exponent. In Fig. \ref{Fig2} the order parameter PDF is plotted in the 
first-scaling form for both models. 

For the XY-model, the three curves exhibit almost perfect collapse. 
Relation (\ref{fS}) is clearly satisfied at $T=0.6$. Similar behavior was observed previously for the XY-model at 
$T_{\mbox{\tiny BKT}}$ \cite{RicBot}. The definite shape of the scaled distribution is 
Weibull-like \cite{FarParBot} similarly to the $T_{\mbox{\tiny BKT}}$ case \cite{RicBot}. 

For the LL model, collapse is not realized in Fig. \ref{Fig2}. As the system size is increased, 
the scaled distributions tend to separate for $T=0.4$. This is evidence that the LL model is {\em not} 
at a critical point for this temperature.

{\it Binder's cumulant check. --} For a continuous phase transition the Binder's cumulant, 
\begin{equation}
U_4 = 1 - \frac{\langle m^4\rangle}{3\langle m^2\rangle}, \label{U4}
\end{equation}
is known to be a universal quantity independent on $L$ at the critical point \cite{Binder}. 

\begin{figure}[h]
\includegraphics[width=8 cm]{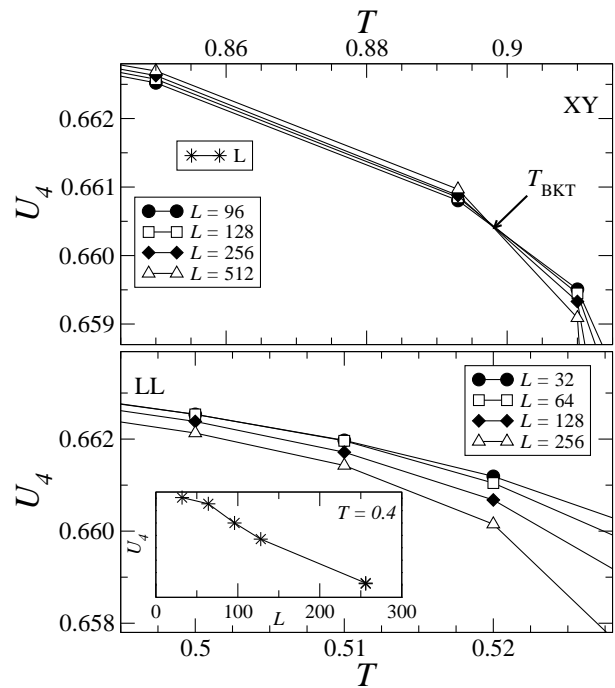}
\caption{\label{Fig3} The Binder cumulant {\it vs} the temperature for the XY-model $T=0.6$ (top) and the LL model (bottom). The number of independent realizations is $10^5$ for each $T$ and $L$. No crossing is observed for the LL model. 
Therefore, no evidence of any phase  transition is observed for the LL model. Values of the Binder cumulant for $T=0.4$ are shown on the inset as a function of the system size. We used $6\times10^6$ independent realizations to obtain each point, so that the error bars are much smaller than the symbol size. }
\end{figure}

For the XY-model, $U_4$ is universal for $T \le T_{\mbox{\tiny BKT}}$ \cite{Loison}. It is checked on the 
Fig. \ref{Fig3} where $U_4$ is plotted for this model (above). 
For the XY-model  a crossing point is observed near the reported BKT temperature. 
For temperatures below the crossing point, the $U_4$ grows with the system size. 
All the curves are expected to collapse in this interval when $L\rightarrow \infty$. 
It is faster when the temperature is small \cite{FarParBot}. On the other hand, 
the $U_4$ above the crossing point, decreases with increasing $L$. 
This type of behavior is observed in others $O(2)$ models with $Z_2$ symmetry \cite{FPB_PRE,BP_CMP}. 

The behavior of $U_4$ is completely different for the LL model. 
The Binder cumulant decreases with $L$ in all the domain of temperature explored ($T>0.1$). 
No crossing  is observed anywhere in Fig. \ref{Fig3} (bottom).

\begin{figure}[h]
\includegraphics[width=8 cm]{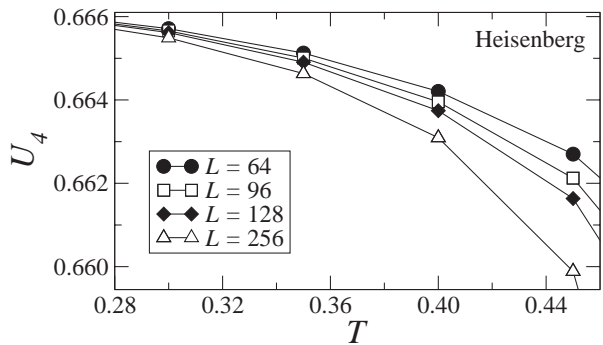}
\caption{\label{Fig4} The 2d Binder cumulant for the Heisenberg model exhibits the same type of behavior as for the LL model (see Fig. \ref{Fig3}).}
\end{figure}

To complement the discussion, we study the Binder cumulant behavior for the Heisenberg model at $T>0.1$. 
It is seen to behave very similarly to the 2d LL model, as no crossing is observed (see Fig. \ref{Fig4}). 
Then we can conclude that in the low temperature range, the LL model  must have very large 
(but not infinite) correlation length, that suddenly begin to decrease in the neighborhood of $T=0.513$. 
For this reason an {\em apparent} QLRO phase may be observed below this temperature.

{\it Discussion. --} We presented in this paper three strong evidences supporting the idea that the 
2d liquid crystals do not have 
a quasi-long-range order phase, namely: \\
(a) the hyperscaling (\ref{hyperscaling}) is not satisfied; \\
(b) the first-scaling collapse (\ref{fS}) does not hold; \\
(c) the Binder cumulant (\ref{U4}) does not exhibit any crossing point.\\
Then this system can not experience a transition of the BKT type.
 
From FSS analysis, Mondal and Roy \cite{Mondal}  concluded that the
LL model should present a continuous transition at $T = 0.548$. The lack
of crossing event for the Binder cumulant behavior (as observed in Fig. \ref{Fig3}) definitely suggests that this is not the case.
In reference \cite{Tesis,FarParBot} the stiffness and the susceptibility are studied as functions of temperature $T$ and system size $L$ for 
the LL- and the XY-models.
For the XY-model the stiffness saturates to finite value below  $T_{\mbox{\tiny{BKT}}}$. However for the LL model the stiffness tends to decrease logarithmically with the system size, similar to the behavior 
of the fully frustrated anti-ferromagnetic Heisenberg model (FFAH) \cite{Wintel}. On the other hand the susceptibility for the LL model changes its functional form in a small region  of temperature 
around $T=0.513$. This is also observed in the FFAH \cite{Wintel}. Then for this reason, and knowing the fact that topological defects are 
stable \cite{Dutta}, we speculate that the LL model may have a crossover 
similar to FFAH.

The set of {\em critical-exponents free} methods used in this article can be used to explore any thermodynamic
systems and to identify possible critical points. The hyperscaling relation 
and the first-scaling law are of great utility to identify 
whether a system is or is not at a critical point. 
In particular, such procedure  could be helpful  for the Heisenberg model in 
the $T<0.1$ domain, to discuss about a possible transition at very low 
temperature \cite{Tesis, Niedemayer,PatrasciouSeiler,Patrasciou, Aguado}. 


AIFS and RP like to thank Bertrand Berche for stimulating discussions about the subject of this article.

\end{document}